\documentclass[]{aastex631}
\usepackage{amsmath}
\usepackage{verbatim}
\usepackage{booktabs}
\usepackage{xcolor}

\shortauthors{Gupta et al.}

\graphicspath{{./}{figures/}}

\def\chandra {{\it Chandra}~}

\def\xmm {{\it XMM-Newton}~}

\def\suzaku {{\it Suzaku}~}
\def\suzakun {{\it Suzaku}}

\def\erosita {{\it eROSITA}~}

\def\mbmxii {{\it MBM12}~}
\def\mbmxiin {{\it MBM12}}

\def\mbmxvi {{\it MBM16}~}
\def\mbmxvin {{\it MBM16}}
\def\mbmxx {{\it MBM20}~}
\def\mbmxxn {{\it MBM20}}
\def\mbmxxxvi {{\it MBM36}~}
\def\mbmxxxvin {{\it MBM36}}

\def\eridanus {{\it Eridanus-hole}~}

\def\g {{\it G236+38}~}
\def\gn {{\it G236+38}}
\def\f {{\it off-filament}~}
\def\fn {{\it off-filament}}

\def\onf {{\it on-filament}~}

\def\oi{{{\rm O}\,{\sc i}~}}

\def\ovii{{{\rm O}\,{\sc vii}~}}

\def\oviii{{{\rm O}\,{\sc viii}~}}

\def\neix{{{\rm Ne}\,{\sc ix}~}}

\def\mgxin{{{\rm Mg}\,{\sc xi}}}
\def\nvi{{{\rm N}\,{\sc vi}~}}

\def\nvii{{{\rm N}\,{\sc vii}~}}

\begin{document}



\title{Where is the Supervirial Gas? III. Insights from X-ray Shadow Observations and a revised Model for the Soft Diffuse X-ray Background}

\correspondingauthor{Anjali Gupta}
\email{agupta1@cscc.edu}

\author{Anjali Gupta}
\affiliation{Columbus State Community College, 550 E Spring St., Columbus, OH 43215, USA} 
\affiliation{Department of Astronomy, The Ohio State University, 140 West 18th Avenue, Columbus, OH 43210, USA} 

\author{Smita Mathur}
\affiliation{Department of Astronomy, The Ohio State University, 140 West 18th Avenue, Columbus, OH 43210, USA} 
\affiliation{Center for Cosmology and Astroparticle Physics, 191 West Woodruff Avenue, Columbus, OH 43210, USA}

\author{Joshua Kingsbury}
\affiliation{Department of Astronomy, The Ohio State University, 140 West 18th Avenue, Columbus, OH 43210, USA} 

\author{Esma Korkmaz}
\affiliation{Columbus State Community College, 550 E Spring St., Columbus, OH 43215, USA} 

\author{Sanskriti Das}
\altaffiliation{Hubble Fellow}
\affiliation{Kavli Institute for Particle Astrophysics \& Cosmology, Stanford University, 452 Lomita Mall, Stanford, CA 94305, USA} 

\author{Yair Krongold}
\affiliation{Instituto de Astronomia, Universidad Nacional Autonoma de Mexico, 04510 Mexico City, Mexico} 

\author{Manami Roy}
\affiliation{Department of Astronomy, The Ohio State University, 140 West 18th Avenue, Columbus, OH 43210, USA} 
\affiliation{Center for Cosmology and Astroparticle Physics, 191 West Woodruff Avenue, Columbus, OH 43210, USA}

\author{Armando Lara-DI}
\affiliation{Instituto de Astronomia, Universidad Nacional Autonoma de Mexico, 04510 Mexico City, Mexico} 

\begin{abstract}

Shadow observations provide a powerful tool to separate foreground components of the soft diffuse X-ray background (SDXB) from the background components. Such observations have now established that the ``local'' foreground is made of the solar wind charge exchange and the local bubble, and the background emission is from the extended circumgalactic medium (CGM) of the Milky Way and from the unresolved extragalactic sources. New data and careful analyses of the SDXB led to two new discoveries in recent years: (1) excess emission near $\rm 0.5 ~keV$ that is identified as the \nvii emission line, and (2) excess emission near $\rm 0.8-1.0~keV$ that is identified with an additional, super-virial temperature hot thermal component of the CGM. The goal of this paper is to use \suzaku shadow observations along six sightlines to determine whether either of these components is from the ``local'' sources. We eliminate the ambiguity regarding the origin of \nvii emission, ruling out the local origin. We confirm that the Milky Way CGM contains nitrogen-rich plasma, with a super-solar average (N/O) of $\rm 2.6\pm0.5$, and suggest that nitrogen-enhanced plasma is widespread throughout the CGM. We find super-solar Ne abundance in two sighlines, also from the CGM. Similarly, we rule out the local origin of the hot thermal component and confirm that it is present beyond the shadowing clouds. Furthermore, we provide a revised model of the soft diffuse X-ray background, which is crucial for extragalactic astronomy.

\end{abstract}

\section{Introduction}
The circumgalactic medium (CGM) is the gaseous halo that surrounds the stellar disk of a spiral galaxy. It is the most extensive and massive baryonic component of a galaxy, playing a crucial role in galaxy formation and evolution. Since most of the CGM is warm-hot, X-ray observations are essential to understand its properties. Our unique vantage point within the Milky Way offers an unparalleled opportunity to study the CGM of a spiral galaxy.

The warm-hot CGM of the Milky Way has been observed in both X-ray absortpion and emission studies \cite[see][for a comprehensive review]{Mathur2022}. We understand that the CGM is extended to over 100 kpc, is diffuse (density $\rm \sim 2.0 \times 10^{-4}~cm^{-3}$), and massive (over $\rm 6 \times 10^{10}~M_{\odot}$) \citep{Gupta2012}. The temperature of this volume-filling component of the CGM was determined to be $\rm \sim 0.20~keV$ (about $2\times 10^6$K), close to the virial temperature of the Milky Way. Hereafter, we refer to this component as the warm-hot phase. While the X-ray observations could not determine the absolute metallicity of the gas, the relative abundance of metals was found to be consistent with solar. 
However, there have been some exciting new developments in our understanding of the Milky Way's CGM, which have challenged previous assumptions about its thermal and chemical structure and offered promising new insights. 

A super-virial temperature hot component at $\rm T \sim 10^7~ K$ with non-solar abundance ratios was first discovered in absorption by \citet{Das2019a}. Several subsequent studies confirmed this result in absorption \citep{Das2021, Lara-DI2023, Lara-DI2024a, Lara-DI2024b, McClain2024, Roy2024b}. Hereafter, we will refer to this super-virial temperature phase as the hot phase.

The discovery of the super-virial hot component was confirmed by emission studies as well \citep{Das2019b, Gupta2021, Gupta2023, Bluem2022, Ponti2023, Bhattacharyya2023}. In addition to the well-known $\rm \sim 0.20~keV$ gas,  $\rm 0.4-1.2~keV$ plasma was discovered. There were also reports of suggestive evidence of nitrogen-rich plasma in the warm-hot CGM of the Milky Way \citep{Das2019a, Gupta2023, Bhattacharyya2023}. 

Thus the presence of the super-virial temperature hot component of the CGM with non-solar abundance ratios has been well established, both in emission and in absorption. However, we also understand that the location, physical characteristics, and the origin of this phase as observed in emission is likely different than that as observed in absorption \citep[see, e.g. discussions in][]{Mathur2022, Bhattacharyya2023}. The focus of this paper is on the emitting gas only. 

\citet{Bisht2024} have argued that the hot emitting gas resides in an extraplanar region beyond the interstellar medium of the Galaxy, and is a result of the stellar feedback. Using idealized simulations based on FIRE, \citet{Roy2024a} also find that this phase resides in the extraplanar region. However, in their model it originates from compressive heating of the gas accreting from the CGM on the disk. The goal of this paper is to determine the location of this phase observationally. This can only be done using a technique called ``shadow observations'' of the soft diffuse X-ray background which we discuss below.

\subsection{Soft Diffuse X-ray Background (SDXB)}

The X-ray emission originating from the Milky Way CGM is derived from observations of the soft diffuse X-ray background  (SDXB), making it imperative that we understand the SDXB. The X-ray background (XRB) above $\rm 1~keV$ exhibits remarkable isotropy on large angular scales and predominantly originates from extragalactic sources. Approximately 90 percent of this background has been successfully resolved into discrete sources \citep[e.g., ][]{Brandt2021}. Below $\rm 1~keV$, the XRB (known as the SDXB) comprises several competing sources of foreground and background components. Traditionally, the SDXB has been modeled using three to four components noted below. 

\begin{enumerate}
  \item A foreground component consisting of Local Hot Bubble (LHB) and Solar Wind Charge eXchange (SWCX):
  \vspace{0.1cm}\\
  The Sun is situated within a low-density region of the local interstellar medium known as the local cavity, with a radius of approximately $\rm 60-200 pc$. This region is believed to be filled with warm-hot ($\rm 10^{6}~K$) plasma and emitting in the soft X-ray band. The plasma within the local cavity is referred to as the LHB and is thought to have been generated by a series of several supernova explosions within the last 10 million years \citep[e.g., ][]{Liu2017, Smith2001}.
  
  SWCX is produced when highly ionized solar wind particles interact with neutral gas, causing an electron to transition from a neutral atom to an excited state in an ion. The electron then cascades to the lower energy level in the ion, emitting soft X-rays and other lines in the process. The SWCX emission is influenced by the properties of the Solar wind, leading to variations in both spectral composition and flux over time. The SWCX spectrum in the $\rm \frac{3}{4}~keV$ ($\rm \sim 0.47-1.21~keV$) range contains lines from various species and ionization states. The most prominent lines are those of \ovii ($\rm 0.57~keV$) and \oviii ($\rm 0.66~keV$ and $\rm 0.81~keV$), with occasional strong contributions from \neix ($\rm 0.92~keV$) and \mgxin ($\rm 1.35~keV$) \citep[][their Fig. 3]{Kuntz2019}.  
    
  Both of these components (LHB and SWCX) are considered ``foreground" components, as they are not absorbed by the cool/cold interstellar medium (ISM) of the Galaxy. Consequently, their spectra are typically modeled as ``unabsorbed" components in the spectral analysis of the SDXB.

  \item A background component made of Cosmic X-ray Background (CXB)
  \vspace{0.1cm}\\
  A significant contribution to the diffuse XRB arises from unresolved sources (e.g., active galactic nuclei, external galaxies, clusters of galaxies). The CXB spectrum is typically modeled as a power-law with a photon index of $\rm \sim 1.45$ and a normalization of about $\rm \sim 10.5~ph~cm^{-2}~s^{-1}~sr^{-1}$ \citep{Cappelluti2017}. As the CXB is extragalactic, it undergoes absorption by the ISM of our Galaxy, thus its spectrum is modeled as an absorbed power-law.
    
  \item The Milky-way CGM or the Galactic Halo
  \vspace{0.1cm}\\
  The emission from the Milky Way's CGM is typically represented by a single-temperature ($\rm \sim 2 \times 10^{6}~K~(\sim 0.2~keV)$) thermal plasma in collisional ionized equilibrium (CIE), such as the APEC model in XSPEC, with solar abundances. This emission is also subject to absorption by the cold gas in the Galactic disk \citep[e.g.,][]{Galeazzi2007, Gupta2009, Henley2015, Liu2017}, and therefore the CGM spectrum is modeled with an absorbed thermal component.

  In summary, the SDXB spectrum is typically described by a model consisting of three to four components: (1) an unabsorbed foreground component associated with the LHB; (2) an optional unabsorbed foreground component arising from the SWCX; (3) an absorbed power-law component representing the CXB; and (4) an absorbed thermal plasma component associated with the Milky Way’s CGM. In this paper, we refer to this model as the {\it standard model} of the SDXB. 
  
  \end{enumerate}

\subsection{Excess emission in the soft diffuse X-ray background}
The large field of view of \xmm and \suzaku provided an unprecedented opportunity to study the SDXB in details. When fitting the SDXB spectra with the aforementioned standard model, excess emissions were discovered in the $\rm \sim 0.4-0.5~keV$ band \citep{Das2019b, Gupta2023} and in the $\rm 0.8-1.0~keV$ band \citep{Mitsuishi2012, Das2019b, Gupta2021, Bluem2022, Bhattacharyya2023, Gupta2023}. 

To fit the excess emission at higher energies, the authors added an additional absorbed thermal component to the SDXB standard model, significantly improving the fit. The best-fit temperatures fall within the range of $\rm 0.4-1.0 ~keV$; thus the excess emission at high energies is identified as emission from gas at super-virial temperature. 

\nvii exhibits a strong transition at $\rm 0.501~keV$; therefore, the excess emission at low energies was identified as that due excess nitrogen in the virial temperature gas in the CGM,  \citep[with an average $\rm N/O = 4.2 \pm 0.2$ solar,][]{Gupta2023}. 

However, there is some ambiguity in the above interpretations. The \neix emission line from SWCX can contaminate the excess emission in the high energy band. Generally, the \neix triplet emission (at $\rm 0.92 ~keV$) from SWCX exhibits comparable effectiveness to the \ovii triplet ($\rm 0.57 ~keV$) in terms of cross-sections and emission probabilities. Nevertheless, due to the lower abundance of $\rm Ne^{9+}$ compared to $\rm O^{7+}$ in the solar wind, the final \neix triplet emission is approximately an order of magnitude weaker than the \ovii triplet emission and contributes nearly half of the \oviii line emission at $\rm 0.66~ keV$ \citep{Kuntz2019}. Indeed, some previous \suzaku studies of SDXB have fitted the excess emission in the energy band of $\rm 0.8-1.1 ~keV$ with an alternative model, positing an overabundance of Ne in the warm-hot gas of the CGM \citep{Mitsuishi2012, Gupta2021}. Similarly, the excess \nvii emission may originate from the LHB or the SWCX.

It is not possible to determine the origin of the hot component and \nvii emission from spectroscopy alone, as the absorbed and unabsorbed foreground components have similar spectral shapes, primarily consisting of X-ray lines from highly ionized metals.  Shadow observations provide a valuable tool for distinguishing between SDXB foreground emission and emissions from the Milky Way CGM and other non-local sources.

\subsection{Shadow observations}
The most challenging part in isolating the X-ray emission from the Milky-way CGM is to model the foreground, whose uncertainties can strongly affect the derived properties of the CGM. The ``shadow observations'' provide a unique technique to disentangle the foreground and background components of SDXB. A typical shadow experiment consists of two observations, one in the direction of a high neutral hydrogen column density ($\rm N_{H}$) region (typically molecular clouds) and the other toward a low $\rm N_{H}$ region a few degrees away. Since the high column absorbs most of the X-rays emitted by distant background sources (i.e., Milky-way CGM and CXB), the spectrum of the high $\rm N_{H}$ field consists mostly of the foreground (SWCX+LB) emission. Contrarily, the spectrum of the low $\rm N_{H}$ field includes foreground as well as a large fraction of the contribution from the distant components. By comparing the spectra from the high- and low-$\rm N_{H}$ regions, we can disentangle the foreground and the distant components, and characterize the spectral properties of X-ray emission of the Milky-way CGM and the foreground with high confidence. Using this technique, it was found that the Milky Way CGM contributes significantly to the SDXB \citep[see, ][for a review]{Henley2015}. The measured temperature of the warm-hot gas in the Milky Way CGM is consistent across different sightlines ($\rm 1.6-2.1 \times 10^{6}~K$); however, the emission measures vary significantly ($\rm 4.5-17.1 \times 10^{-3}~cm^{-6}~pc$).

In this paper we present \suzaku shadow observations along six sightlines. Our goal is to understand whether the excess emission in both low- and high-energy bands is from the local sources (SWCX and LB) or from regions beyond the molecular clouds. We present the data in \S 2 and the spectral analysis is presented in \S 3. We rule out the local origin of the excess emission, which is discussed in \S 4. We also present a revised spectral model for the SDXB incorporating the models for the excess emissions. 

\section{Suzaku Shadow Observations and Data Reduction}
\suzaku performed six shadow experiments during $\rm 2006-2013$. Each shadow experiment consists of two nearby observations of low- and high-$\rm N_{H}$ regions. The targets of these six shadow experiments were \mbmxii on- and off-cloud \citep{Smith2007}, on- and off-filament in the Southern Galactic Hemisphere \citep[SGH; ][]{Henley2007}, \mbmxx and \eridanus \citep{Gupta2009}, \g on- and off-cloud \citep{Henley2015}, \mbmxvi on- and off-cloud \citep{Liu2017}, and \mbmxxxvi on- and off-cloud \citep{Ursino2016}. The exact distances of these clouds are uncertain, but they lie between $\rm 60-275~pc$ (Table 1). Our goal is to use these clouds to separate foreground components such as the LHB and SWCX from more distant components such as the Galactic halo emission, to determine the location of the excess emission. Therefore, the small distance uncertainty is not relevant. 

The \suzaku shadow observation pair IDs, dates, pointing directions, and exposure times are listed in Table 1. Fig. 1 shows the locations of these shadow observations on the sky map. The sightline of \mbmxxxvi passes through the Galactic \erosita bubble \citep{Predehl2020}. The sightlines of \mbmxii and \mbmxvi are closely spaced, and the \mbmxx sightline passes through a region near the Eridanus enhancement. 

\subsection{Suzaku Data Reduction}
We performed the \suzaku data reduction using HEAsoft version 6.30. For this study, we focused on data from the back-illuminated (BI) XIS1 detector, as it has greater sensitivity at low energies compared to the front-illuminated (FI) XIS0 and XIS3 cameras. We combined data collected in the $\rm 3 \times 3$ and $\rm 5 \times 5$ observation modes. In addition to the standard screening procedures described in the \suzaku Data Reduction Guide, we applied extra screening to the data. To minimize detector background noise, we excluded times when the cut-off rigidity (COR) of the Earth's magnetic field was less than 8 GV (the default value is $\rm COR > 2 ~GV$). Additionally, we expanded the filter value for the angle between \suzakun's line-of-sight and the limb of the Earth (ELV) from the default 5$^{\circ}$ to 10$^{\circ}$. This adjustment helps remove excess events in the 0.5-0.6 keV band caused by solar X-rays scattered off the Earth's atmosphere \citep{Smith2007}.

One standard screening criterion for \suzaku data involves selecting times when the angle between \suzakun's line-of-sight and the limb of the sunlit Earth ($\rm DYE\_ELV$) is greater than $\rm > 20^{\circ}$. However, \citet{Sekiya2014} noted that \suzaku observations, particularly after 2011, were affected by increased solar activity. The enhanced interaction of solar X-rays with neutral oxygen in the Earth's atmosphere creates an \oi fluorescent line at $\rm E = 0.525~keV$. This \oi emission can significantly contaminate the \nvii and \ovii K$\alpha$ emissions at $\rm 0.501~eV$ and $\rm 0.575~eV$, respectively. These lines cannot be resolved because of the low energy resolution of \suzaku ($\rm \approx 50~ eV$). Indeed, in \citet{Gupta2021} and \citet{Das2019c}, we found that in the \suzaku spectra taken in 2014, the \oi intensity ranged from 25\% to 130\% of the \ovii intensity. Therefore, it is crucial to reduce \oi contamination to accurately constrain the \nvii and \ovii K$\alpha$ emissions, which are major components of the foreground and the Milky Way CGM emission.

\citet{Sekiya2014} suggested choosing events taken during intervals when $\rm DYE\_ELV > 60^{\circ}$ to minimize \oi contamination. However, this results in the loss of a large amount of data. Of the six \suzaku shadow experiments, three (\mbmxvin, \mbmxxxvin, and \gn) were conducted during or after 2011. For these observations, we carefully investigate different limits on the $\rm DYE\_ELV$ parameter, following \citet{Gupta2021}, and select a value that significantly reduces \oi emission while maintaining a good balance between exposure time loss and \oi contamination. For other observations taken before 2011, we applied the standard screening of $\rm DYE\_ELV > 20^{\circ}$.

Since our goal is to analyze the diffuse emission, it is essential to remove point sources. We first generated $\rm 0.5-2.0 ~keV$ images to identify the bright point sources.  The point source exclusion regions, ranging from $\rm 1'-3'$, were selected manually. The spectrum of the diffuse background was extracted from the entire field of view after removing the point sources. 

We produced the redistribution matrix files (RMFs) using the \textit{xisrmfgen} tool, which accounts for the degradation of energy resolution and its position dependence. We also prepared ancillary response files (ARFs) using the \textit{xissimarfgen} tool with the revised procedure\footnote{https://heasarc.gsfc.nasa.gov/docs/suzaku/analysis/xisnxbnew.html}. For the ARF calculations, we assumed a uniform source with a radius of $\rm 20\arcmin~$ and used a detector mask that excluded the bad pixel regions. The total instrumental background was estimated from the night Earth data database using the \textit{xisnxbgen} tool. 

\section{Spectral Analysis}
We used the XSPEC version 12.10.1f \footnote{https://heasarc.gsfc.nasa.gov/xanadu/xspec/} for the spectral fitting. 
We modeled all the thermal plasma components in CIE with the APEC or VAPEC (version 3.0.9) model \citep{Smith2001} and used solar relative metal abundances of \citet{Anders1989}. For absorption by the Galactic disk, we used the {\it phabs} model in XSPEC. 

We used the $IRAS\ 100\micron$ maps to estimate the neutral hydrogen column density ($\rm N_{H}$) across all regions analyzed in this study, following the method outlined by \citet{Galeazzi2007}. By applying the standard high-latitude $IRAS\ 100\micron/N_{H}$ ratio of $\rm 0.85 \times 10^{-20}~cm^{2}MJysr^{-1}$ \citep{Boulanger1988}, we calculated the neutral hydrogen densities for each field. The $\rm N_{H}$ values obtained through this method are listed in Table 1. While this approach provides reliable measurements for regions with relatively low column densities, it may have some uncertainties for denser regions, such as molecular clouds, as noted by \citet{Ursino2016}. Due to these uncertainties, particularly for on-cloud column densities, we treated $\rm N_{H}$ as a free parameter in the fitting process. The resulting best-fit $\rm N_{H}$ values are also presented in Table 1.

The best-fit $\rm N_{H}$ values are consistent with those derived from the $IRAS\ 100\micron$ maps for all sightlines except \mbmxxxvin. We attempted to fit the on- and off-cloud spectra for \mbmxxxvi using $\rm N_{H}$ values fixed to those calculated from the $IRAS\ 100\micron$ data. The resulting fits were poor ($\rm \chi^{2}/dof = 856.73/522$), making it necessary to keep $\rm N_{H}$ as a free parameter. We note that our best-fit $\rm N_{H}$ values for \mbmxxxvi are consistent with those reported by \citet{Ursino2016}.

\subsection{CXB Modeling in Spectral Fitting}
Before fitting the SDXB in the $\rm < 1~keV$ range, we first estimated the CXB, which dominates above $\rm 1.5~keV$. We first simultaneously fitted the spectra from both low- and high-$\rm N_H$ regions in the $\rm 1.5-5.0~keV$ energy range using an absorbed power-law model. The absorption column density was fixed to the measured values. Additionally, the power-law parameters were linked between the on- and off-cloud regions.

This model provided a good fit for the directions \mbmxvin, \mbmxxn, \mbmxxxvin, and \gn, where the CXB is dominant above $\rm 1.5~keV$.  The best-fit CXB parameters for these directions were photon spectral index $\Gamma = 1.39\pm0.07$, $~1.39\pm0.07$, $~1.41\pm0.07$, and $~1.43\pm0.05$, and the normalizations of $9.91\pm0.72$, $8.85\pm0.58$, $10.38\pm0.75$, and $10.38\pm0.38~ \rm photons~keV^{-1}s^{-1}~sr^{-1}~cm^{-2}$, respectively.

However, the above simple model did not fit well for the regions \mbmxiin, and the filament in the SGH. For the on-cloud observation of \mbmxii, which includes the bright source XY Ari, we accounted for contamination in the $\rm 0.3-1.0~keV$ range. Despite XY Ari’s expected lack of emission below 1~keV due to heavy absorption \citep{Salinas2004}, its CCD response tail \citep{Smith2007} contributed to the spectrum. We excluded a 2 arcmin region around XY Ari but found that scattered emission still affected the fit. To address this, we added an additional power-law component (best fit $\rm Gamma = -1.41\pm0.41$, normalization of $0.29\pm 0.17~\rm photons~keV^{-1}s^{-1}~sr^{-1}~cm^{-2}$), which significantly improved the fit. The best-fit CXB parameters for this region were $\Gamma = 1.38\pm 0.11$ and normalization $8.01\pm0.82~\rm photons~keV^{-1}s^{-1}~sr^{-1}~cm^{-2}$.

A similar issue arose in the on-filament observation, where the simple power-law model did not provide a good fit in the $\rm 1.5-5.0~keV$ range. A very bright source within the field of view was contributing additional emission. We added another power-law component (best fit $\rm Gamma = 1.08\pm0.38$, normalization of $2.55\pm 1.01~\rm photons~keV^{-1}s^{-1}~sr^{-1}~cm^{-2}$) to model this contribution, which led to a much better fit. The best-fit CXB parameters in this case were $\Gamma = 1.69\pm 0.12$ and normalization $10.28\pm1.12~\rm photons~keV^{-1}s^{-1}~sr^{-1}~cm^{-2}$.

Using deep \chandra observations of the COSMOS field, \cite{Cappelluti2017} provided one of the most precise measurements of the CXB, reporting an extragalactic power-law spectrum with a photon index of $\Gamma = 1.45 \pm 0.02$ and a 1~keV normalization of $10.91 \pm 0.16\rm ~photons~keV^{-1}~s^{-1}~sr^{-1}~cm^{-2}$. The power-law normalizations derived from our data are consistent with those obtained in the COSMOS field as well as with previous measurements of the SDXB from \xmm and \suzaku observations \citep{Galeazzi2009, Gupta2009}.

\subsection{Modeling the SDXB}
For each pair of shadow observations, we simultaneously fitted the spectra from the low- and high-$\rm N_{H}$ regions. We started by applying the standard model discussed in §1.1, linking the model parameters for the foreground (LB+SWCX) and the background (Galactic halo + CXB) components. For off-cloud regions, the absorption column density was frozen at the measured values, whereas for on-cloud regions, it was allowed to vary as a free parameter. 

We used a two-component model to represent the foreground emission from the LHB and the SWCX. The LHB emission was modeled using a single-temperature thermal APEC model with solar abundances. \cite{Liu2017}, using data from the {\it ROSAT All-Sky Survey} and {\it the Diffuse X-rays from the Local Galaxy} (DXL) sounding rocket mission, produced thermal emission maps for the LHB. We fixed the APEC temperature and normalization to the values obtained from these maps.

The SWCX component was modeled using the ACX2 (v2.2.0) model \citep{Smith2014, Foster2020}, which incorporates velocity-dependent effects and charge exchange cross sections from the Kronos database \citep{Mullen2016, Mullen2017, Cumbee2018}. To simplify the model, we assumed all solar wind ions travel at a uniform velocity equal to the mean solar wind speed ($\rm 450~km~s^{-1}$). The neutral helium fraction was fixed to the cosmic value of 0.09 (default), with solar abundances, single recombination type, and an {\it acxmodel} parameter value of 8. We confirmed that the selection of solar wind speed and the acxmodel parameter has little impact on the shape of the SWCX component for \suzaku XIS spectral resolution.

To model the Galactic halo emission, we started with an absorbed single-temperature thermal emission model (APEC) assuming solar abundances.

For observations obtained during or after 2011—specifically, \mbmxvin, \mbmxxxvin, and \gn, we included a Gaussian line to account for the \oi contamination. First, we fitted the standard model to spectra filtered for $\rm DYE\_ELV > 20^{\circ}$, $\rm DYE\_ELV > 40^{\circ}$, and $\rm DYE\_ELV > 60^{\circ}$, and measured the \oi line normalization for each.

For \mbmxvi, we found that the \oi line strength decreased from $\rm 0.98\pm0.36~LU$ to a value consistent with zero ($\rm 0.22\pm0.31$) for $\rm DYE\_ELV > 20^{\circ}$ to $\rm DYE\_ELV > 40^{\circ}$, with no further change for $\rm DYE\_ELV > 60^{\circ}$. Based on this, we used the $\rm DYE\_ELV > 40^{\circ}$ filter for further analysis of \mbmxvi.

For \mbmxxxvi and \gn, the \oi line strength is consistent with zero, and we found no significant difference in \oi normalization across the three elevation angle thresholds. Therefore, we adopted the $\rm DYE\_ELV > 20^{\circ}$ filter for further spectral fitting of both \mbmxxxvi\ and \gn.

We note that for all pairs of the shadow observations the on-cloud high-$\rm N_{H}$ spectrum has significantly lower flux than the off-cloud low-$\rm N_{H}$ spectrum. This confirms that the SDXB emission is not dominated by local sources (LB$+$SWCX), but is mostly from the regions beyond the absorbing disk \cite[Milky Way CGM $+$ CXB,][]{Gupta2009b}.

\subsubsection{\mbmxxxvi}
The standard model discussed above provided a good fit to the \mbmxxxvi on-cloud high-$\rm N_{H}$ spectra ($\rm \chi^{2}/dof = 254.85/214$), but a poor fit ($\rm \chi^{2}/dof = 406.40/312$) to the off-cloud low-$\rm N_{H}$ spectra showing excess emissions at both low ($\rm \sim 0.4-0.5~keV$) and high ($\rm \sim 0.8-1.0~keV$) energy bands. The best fit statistics of combined simultaneous fit is $\rm \chi^{2}/dof = 661.25/526$. Figure 2 shows the spectral fits to the \mbmxxxvi on- and off-cloud regions. As discussed in \S 1.2, similar excess emissions at these energy bands have been reported in previous studies. Careful comparisons of the on-cloud and off-cloud spectra clearly shows that the excess emissions at both low and high energies are more pronounced toward the low-$\rm N_{H}$ off-cloud sightlines. This shows that they do not originate from local sources (LB+SWCX), but are from distant regions absorbed by the Galactic disk (e.g. Milky Way CGM).

To fit the excess emission in the higher energy band ($\rm 0.8-1.0~keV$), and following the approach used in previous studies \citep{Gupta2021, Gupta2023}, we introduced an additional absorbed thermal component. Including this component significantly improved the fit ($\rm \Delta \chi^{2}/\Delta dof = 62.81/2$). The best-fit temperature of the added thermal component is $\rm 0.692 \pm 0.031~keV$. This corresponds to the super-virial temperature plasma discussed in \S 1.2. 

We then attempted to model the excess emission in the lower energy band. \nvii has a strong transition at $\rm 0.5~keV$ and the ionization fraction of \nvii peaks at a temperature around $\rm \sim 0.2~keV$. Following our previous studies, we allowed the nitrogen abundance to vary in the absorbed thermal component ($\rm kT \sim 0.2~keV$) corresponding to the Milky Way CGM, to fit the low-energy excess emission. This approach significantly improved the fit ($\rm \Delta \chi^{2}/\Delta dof = 7.84/1$). The excess N/O ratio is required at a significance level of $>99.1\%$ (F-test). 

As noted above, the excess low-energy emission does not originate from local sources. Nevertheless, we performed an additional test to confirm this. We allowed the nitrogen abundance to vary in the foreground (unabsorbed) SWCX component (VACX2), instead of in the absorbed component. This did not improve the fit, $\rm \chi^{2}/dof = 598.21/523$, further supporting the model with enhanced nitrogen in the absorbed thermal component.

Residual features remained in the higher energy band ($\rm 0.8-1.0~keV$). Given that \neix has a transition at $\rm 0.91~keV$, we allowed the neon abundance in the absorbed warm-hot component to vary, which further improved the fit ($\rm \Delta \chi^{2}/\Delta dof = 9.63/1$). As noted for several sightlines in \citet{Gupta2023}, the sightlines toward \mbmxxxvi required both an overabundance of neon in the warm-hot component and an additional hot absorbed thermal component. The best-fit values are reported in Table 2, and Figure 3 shows the best-fit model for the \mbmxxxvi on- and off-cloud regions.

In the spectrum for the \mbmxxxvi off-cloud region, there is also a line-like feature around $\rm 0.40-0.45~keV$. \nvi has a strong transition at $\rm 0.43~keV$. To model this feature, we added a Gaussian line. The best-fit energy is $\rm 0.42 \pm 0.01~keV$ ($\rm \chi^{2}/dof = 572.65/520$). The line strength (normalization) is consistently higher in the low-$\rm N_H$ off-cloud observations compared to the on-cloud observation, again suggesting an origin beyond the absorbing clouds. The detection of \nvi indicates the presence of a low-temperature plasma.


In the spectral analysis above we have used the solar elemental abundances from \citet{Anders1989}. To test whether our results depend on this choice, we further used different abundances from  \cite{Lodders2009, Asplund2009, Lodders2003, Wilms2000}. With these different abundance values, the on-cloud best-fit absorption column density increased to higher values in the range $\rm 6.3-6.9 \times 10^{20}~cm^{-2}$ from $\rm \sim 4.6 \times 10^{20}~cm^{-2}$ with the \citet{Anders1989} abundances. However, our results on the Milky Way CGM were not affected by the choice of the solar abundance model. The hot super-virial temperature component was still required as did the super-solar N/O and Ne/O abundances.

Although the \mbmxxxvi observations were taken after 2011, as discussed above, we adopted a filter of $\rm DYE\_ELV > 20^{\circ}$ for spectral fitting. To ensure consistency, we also fit our best-fit model to data filtered with $\rm DYE\_ELV > 60^{\circ}$. The resulting model parameters remain consistent with those obtained using the $\rm DYE\_ELV > 20^{\circ}$ threshold. 

\subsubsection{Filament in the SGH}
Simultaneous fitting of the \f and \onf spectra in the SGH resulted in a poor fit for both fields ($\rm \chi^{2}/dof = 758.94/676$). The SWCX component from the ACX2 model was unconstrained and showed negligible emission. In the standard model, we assumed that the SWCX remained constant between the on- and off-cloud observations. However, it is possible that the SWCX varied between the two, even though the observations were taken only two days apart.

To test this, we allowed the SWCX parameters to vary between the on- and off-filament spectra. This did not improve the fit, and the SWCX emission remained negligible in both cases.

Thus, for the filament shadow observations, we modeled the foreground component using only an APEC model. When the model parameters were tied together, the emission measure (EM) of the foreground APEC was poorly constrained, with a fit  statistic of $\rm \chi^{2}/dof = 737.60/676$. We then fixed the temperature to $0.1~\rm keV$ and allowed the EM of the unabsorbed APEC component to vary between the on- and off-cloud fields. This significantly improved the fit ($\rm \chi^{2}/dof = 714.88/676$), and the APEC parameters became well constrained.

Excess emission in the higher energy band ($0.8$–$1.0~\rm keV$) is clearly visible, though the low-energy excess is less apparent by eye. To account for the high-energy excess, we added another thermal component, which significantly improved the fit ($\rm \chi^{2}/dof = 692.07/674$). This additional thermal component is required at more than 99.998\% confidence (F-test). 

Although the low-energy excess is not as strong as in \mbmxxxvin, some excess is still present toward \fn. To model this, we allowed the N abundance in the warm-hot component to vary, which led to a marginal improvement in the fit ($\rm \chi^{2}/dof = 687.87/673$), with a best-fit value of $\rm N/O = 1.6 \pm 0.3$. This enhanced N/O ratio is required at a significance level of 95.69\%. 

Subsequently, allowing the Ne abundance in the warm-hot component to vary led to a further improvement in the fit ($\rm \chi^{2}/dof = 675.02/672$). The overabundance of Ne is required at significance at more than 99.96\%. The best fit parameters are reported in Table 2.

\subsubsection{\mbmxii}
The simultaneous fitting of the standard model provided a good fit to the \mbmxii on-cloud observations. However, in the off-cloud field, we observed excess emission at higher energies, near $\sim0.85~\rm keV$. Adding another thermal component to the model did not explain this extra emission and did not improve the fit.

\citet{Smith2007} first analyzed data from \mbmxii and noticed similar extra emission between $0.8$ and $0.9~\rm keV$. They added a Gaussian line at $E = 0.876~\rm keV$ to account for it and called it a ``mystery feature'' since no known atomic lines match this energy. Using the same approach, we added a Gaussian line to our model and found a similar feature at $E = 0.859 \pm 0.016~\rm keV$ with an intensity of $0.24 \pm 0.11~\rm LU$. The cause of this feature is still unknown, but future missions like ATHENA, which will have better resolution, might help us understand it.

To check for overabundance of N/O, we allowed the nitrogen abundance to vary. This slightly improved the fit ($\Delta\chi^2/\Delta\mathrm{dof} = \rm 3.77/1$, F-test probability of $>93.6\%$), with $\rm N/O = 3.1\pm1.3$ (Table 2).

\subsubsection{\mbmxx}
 The simultaneous fit using the standard model provided a good fit ($\chi^2/dof = \rm 649.03/623$), however, the ACX2 normalization was not well constrained. There was also some extra emission at lower energies in the off-cloud spectrum. To fit the excess emission at lower energies, we required an overabundance of $\rm N/O = 3.7\pm1.2$, with an F-test probability of $>99\% ~$ (Table 2). An extra thermal component was not needed.

Since the ACX2 model was poorly constrained for the \mbmxx direction, we also tested a model using only APEC for the foreground emission. In this case, the temperature and normalization parameters were linked between the on- and off-cloud observations, and were allowed to vary. The fit statistics and the parameters for the Milky Way’s CGM remained largely unchanged.

\subsubsection{\mbmxvi}
The standard model provided a good fit to both on- and off-cloud observations toward \mbmxvi (Table 2). No additional thermal component or excess N/O was statistically required.

\subsubsection{\g}
The standard model provided a good fit to both on- and off-cloud observations towards \g\ ($\chi^2/\mathrm{dof} = \rm 457.20/447$), though slight excess emissions were noted at both low and high energy bands. Introducing a variable N/O ratio resulted in a small improvement in the fit ($\chi^2/\mathrm{dof} = \rm 455.78/453$, Ftest $P > 76\%$). Additional thermal component is not constrained.

\textbf{In summary, two (\mbmxxxvi\ and \mbmxxn) out of six sightlines show a significant overabundance of N/O in the warm-hot component of the Milky Way CGM, with a confidence level greater than 99\%. Additional two ({\it Filament} and \mbmxiin) sightlines exhibit marginal detections of N/O enhancement (at $> 95\%$ and $> 93\%$ respectively). Furthermore, the sightlines toward \mbmxxxvi\ and {\it Filament} also show an overabundance of Ne/O in the warm-hot component and require an additional, hotter plasma component, both detected at high significance.} 


\section{Discussion}
\subsection{Alternative Models}
Several previous studies have employed different models to characterize the SDXB or Milky Way’s CGM emission. For example, \cite{Miller2008} used a variable abundance model to fit the \suzaku spectrum of the North Polar Spur (NPS). Notably, the shadow sightline of \mbmxxxvi lies close to the NPS. \cite{Kataoka2021} analyzed \suzaku observations of the SDXB and reported a best-fit CGM temperature of $\rm 0.3~keV$ with sub-solar metallicity ($Z = 0.2Z_{\odot}$). More recently, \cite{Yamamoto2022} investigated plasma conditions in the NPS/Loop I region and proposed a non-equilibrium ionization (NEI) state. Additionally, \cite{Gu2016} suggested that ionized absorption could account for the \suzaku data from the NPS. Motivated by these studies, we tested these models to assess whether our two-temperature model—with enhanced N/O and/or Ne/O abundance ratios—provides a better fit.

Among the six sightlines in our sample, the \mbmxxxvi observation has the highest signal-to-noise ratio (S/N), making it most suitable for testing alternative models. The variable abundance model of Miller et al. yields an improved fit over the standard model, with fit statistics of $\chi^2/\mathrm{dof} = \rm 612.42/521$. Similarly, the sub-solar abundance model by Kataoka et al. ($\chi^2/\mathrm{dof} = \rm 644.04/526$) and the NEI model by Yamamoto et al. ($\chi^2/\mathrm{dof} = \rm 618.49/525$) also improve upon the standard model fit ($\chi^2/\mathrm{dof} = \rm 661.25/526$).

We tested the ionized absorber scenario using the ISMabs model in XSPEC \citep{Gatuzz2015}, but it yielded a poor fit ($\chi^2/\mathrm{dof} = \rm 700.51/525$). Allowing the \oi abundance to vary led to a modest improvement ($\chi^2/\mathrm{dof} = \rm 666.61/524$). Further allowing the abundances of N, O, Ne, Mg, and Fe to vary \cite[following][]{Gu2016} significantly improved the fit ($\chi^2/\mathrm{dof} = \rm 636.08/519$).

Our best-fit model for \mbmxxxvi, with supersolar (N/O) and (Ne/O) and a super-virial temperature component, with fit statistics of $\chi^2/\mathrm{dof} = \rm 572.65/520$, provides a significantly better fit than all other models discussed above.

\subsection{Enhanced nitrogen abundance}
We detect excess emission near $\rm 0.5~keV$, which is well-fit by a \nvii\ emission line. This feature is significantly stronger along off-cloud sightlines, supporting the conclusion of \citet{Gupta2023} that the emission arises from an enhanced N/O abundance in the CGM of the Milky Way, rather than from SWCX. In the present work, we detect nitrogen-rich plasma with high significance in two sightlines and marginally in two others. The sightlines with marginal or non detection have fainter emission, limiting our ability to constrain the N/O ratio. A similar trend was reported in \citet{Gupta2023}, where enhanced N/O was detected at high significance in brighter sightlines (both within and outside the Galactic bubbles), and only marginally in fainter ones.  Based on results of this paper and those of \citet{Gupta2023}, we suggest that nitrogen-enhanced plasma is widespread in the CGM, and that non-detections are primarily due to limited signal-to-noise, rather than indicating the absence of enriched plasma. A more comprehensive study of the nitrogen distribution across the sky will be presented in Gupta et al. (2025, in preparation).

The observed nitrogen overabundance can be understood in the context of stellar nucleosynthesis. According to the standard model of stellar metal enrichment, carbon and oxygen are predominantly produced in massive stars via helium shell burning and are released into the interstellar medium (ISM) through supernova explosions. In contrast, nitrogen is primarily synthesized in intermediate-mass stars (4–8 $M_\odot$) and expelled via winds during the asymptotic giant branch (AGB) phase \citep{Henry2000, Miller2008}. Although evolved high-mass stars such as Wolf–Rayet stars and Luminous Blue Variables can also contribute to nitrogen production, their role is comparatively minor \citep{Henry2000}. The marked nitrogen enhancement observed in the CGM thus points toward significant enrichment from AGB stellar feedback.

Given that the CGM acts as a long-term reservoir that retains and mixes metals from multiple generations of star formation, the widespread N/O enhancement implies that feedback driven by AGB stars has played a sustained and dominant role in shaping the CGM’s chemical composition. The widespread enhancement of the N/O across the sky further suggests efficient mixing on large scales, potentially driven by mechanisms such as galactic fountains, slow winds, or turbulent diffusion, rather than being solely the result of localized star formation or AGN-driven outflows.

\subsection{The hot component}
To better fit the data in the higher energy band of $\rm 0.8-1.0~keV$ along two sightlines, we needed an additional absorbed thermal component at temperatures in the range of $\rm 0.8 - 0.75~keV$.  Once again, the excess emission at the higher energy band is more pronounced toward off-cloud sightlines, indicating that this is due to an absorbed thermal component rather than \neix emission from SWCX.

Among the two sightlines where we have observed the hot component, one sightline towards \mbmxxxvi passes through bright structures of the Galactic bubbles near the NPS, while the other sightline probes the region outside the bubbles. In an all-sky survey using \suzaku single-field observations, \citet{Gupta2023} reported that the super-hot component is present both inside and outside the bubble region, although the EM is notably higher towards sightlines passing through the bubbles. In our shadow observations, we have observed the same phenomenon.

We emphasize that the detection of the hot component along two out of six sightlines does not imply that the covering fraction of this component is $33\%$. The four sightlines where the hot component was not detected exhibit an order of magnitude fainter emission compared to the average sky emission. According to \citet{Gupta2023}, the EM of the warm-hot component, excluding the bubbles region, ranges from $\rm 0.8-14.2 \times 10^{-3}~cm^{-6}~pc$, with an average of $\rm 4.4 \times 10^{-3}~cm^{-6}~pc$. In contrast, the sightlines where no hot component was detected show EM values for the warm-hot component ranging from $\rm 1.63-3.23 \times 10^{-3}~cm^{-6}~pc$. This lower EM resulted in low S/N spectra, making it difficult to detect the even fainter hot component.

In this work we show that the hot component resides beyond the absorbing Galactic disk, though its exact location cannot be determined by shadow observations. As noted in \S 1.2, \citet{Bisht2024} and \citet{Roy2024a} argue that the hot gas resides in an extra-planar region outside of the stellar disk. Our results are consistent with these models. 

\subsection{Enhanced neon abundance}
In the directions of \mbmxxxvi and the SGH filament, we detect not only a hot gas component but also evidence for a supersolar Ne/O abundance ratio in the warm-hot phase. Gupta et al. (2023) reported similar findings along ten sightlines passing through the  Galactic bubbles, where the best-fit models required an average Ne/O ratio of $2.1 \pm 0.2~$ solar. Our measurements are consistent with these results; however, the filament sightline analyzed here is located outside the regions associated with Galactic bubbles. This indicates that the supersolar Ne/O enhancement may not be limited to the bubble interiors, but could instead represent a more widespread characteristic of the Milky-way CGM.

\subsection{Revised model of the SDXB}
As noted above (\S 1.1), the SDXB is typically modeled with a three-four component model: one or two unabsorbed foreground components (LHB+SWCX) and two absorbed components (the warm-hot Milky Way CGM and the CXB). However, this work, in agreement with recent studies, confirms that this model is insufficient to characterize  the SDXB spectrum. This is not only important for the Galactic halo studies but is also crucial for the investigation of extended sources such as supernova remnants, galaxies, galaxy clusters, and the intergalactic medium, where the source of interest fills the entire field-of-view. In this section, we summarize the updated model of the SDXB. 

\begin{enumerate}
  \item A foreground (Unabsorbed) component of LHB: The LHB emission is well-described by a single-temperature thermal model with solar abundances. \cite{Liu2017} present the temperature and EM maps for the LHB.
  \item A foreground (Unabsorbed) component of SWCX: The SWCX emission can be modeled using the ACX2 model. 
  \item A background component of CXB:
    There are no proposed changes on how to model the CXB. The CXB spectrum is well characterized by an absorbed power law with a photon index close to $\rm 1.45$. In both this study and our previous ones, we have permitted variations in the photon index and normalization for improved fit results.
  \item The warm-hot Component of the MW CGM:
    Typically, this component is characterized by a constant-abundance (with respect to solar) thermal plasma model. However, in our case, we are needed to model this component using a variable abundance model (e.g., vapec in XSPEC), allowing to vary abundances of nitrogen (N), and neon (Ne). This change significantly improves the fit.
  \item The hot Component of the MW CGM:
    This is the new component required to model emission at higher energy band of $\rm 0.8-1.0~keV$. A single-temperature thermal model with constant abundances are sufficient to model this component in CCD resolution data. 
   \item Low energy line like emission:
    In a few datasets, there is a line-like feature at approximately $\sim 0.4~\rm keV$. This feature can be modeled by adding a Gaussian line.
\end{enumerate}

\section{Conclusion}
In this paper we use the unique power of shadow observations to disentangle foreground and background components of the SDXB. We present a revised spectral model for the SDXB. In addition to the three-component model made of SWCX$+$LB, warm-hot CGM of the Milky way and the CXB, we show that the following components are necessary. (1) Hot component from beyond the Galactic disk; (2) excess amount of \nvii emission; (3) excess amount of \nvi emission, possibly from a lower temperature plasma; and (4) excess amount of \neix emission. Both, super-solar N and Ne are in the warm-hot component of the Milky Way CGM. We conclusively show that the hot component is from beyond the Galactic disk, possibly from the extraplanar region, as suggested by recent theoretical models. Note that the Milky Way CGM results presented here are for the X-ray emitting component of the CGM. The diffuse, extended, and massive component of the CGM, detected in X-ray absorption, is not the subject of this work. 

\section{Acknowledgements}
We thank the anonymous referee for insightful comments that improved the clarity and robustness of this work. We gratefully acknowledge support from NASA ADAP grants 80NSSC24K0626 and 80NSSC22K0480 to AG. AG and SM also acknowledge support from the National Aeronautics and Space Administration (NASA) through Chandra Award Numbers GO3-24126X and AR0-23014X, respectively, issued by the Chandra X-ray Center, which is operated by the Smithsonian Astrophysical Observatory on behalf of NASA under contract NAS8-03060. SM additionally acknowledges support from NASA ADAP grant 80NSSC22K1121. SD acknowledges support from the NASA Hubble Fellowship and the KIPAC Fellowship of the Kavli Institute for Particle Astrophysics and Cosmology, Stanford University. YK acknowledges support from grant PAPIIT-UNAM IN102023.

\clearpage


\begin{figure}
\includegraphics[scale=1.2]{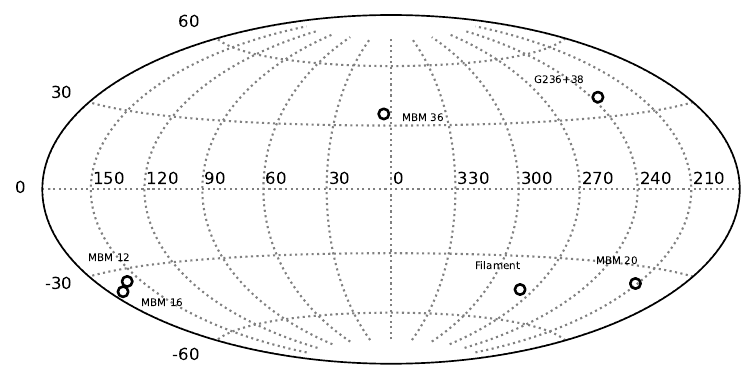}
\vspace*{-0.3 cm}
\caption{All-sky map centered on the Galactic Center, showing the locations of the shadows analyzed in this paper. }
\end{figure}

\begin{figure}
\includegraphics[scale=1.0]{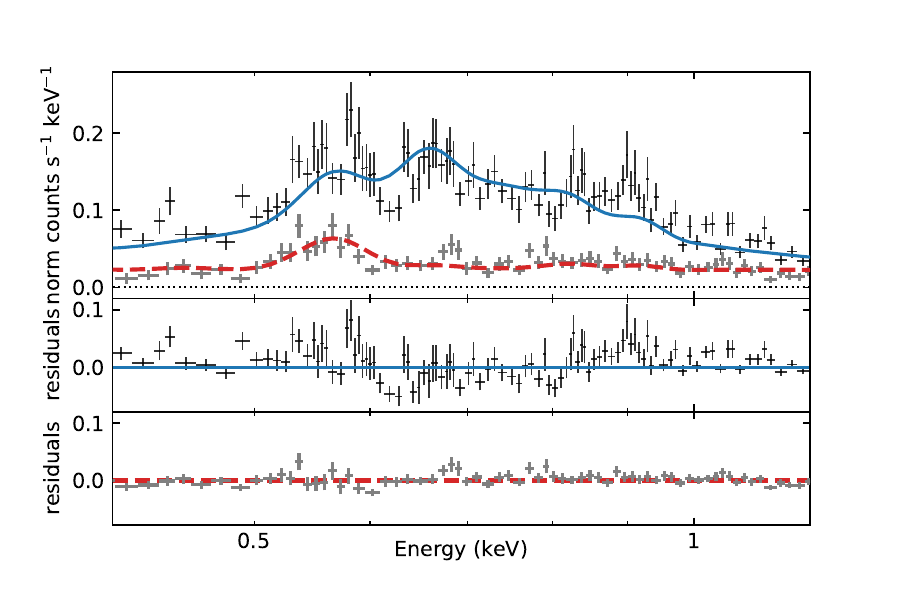}
\vspace*{-0.3 cm}
\caption{\suzaku XIS1 spectra towards MBM36 on-cloud (grey) and off-cloud (black) were fitted simultaneously with the SDXB standard model. Excess emissions at the lower energy and higher energy bands can be clearly seen in the residual plot of the off-cloud spectrum (middle panel). In contrast, excess emissions are not as prominent in the on-cloud spectrum (bottom panel).}
\end{figure}

\begin{figure}
\includegraphics[scale=1.0]{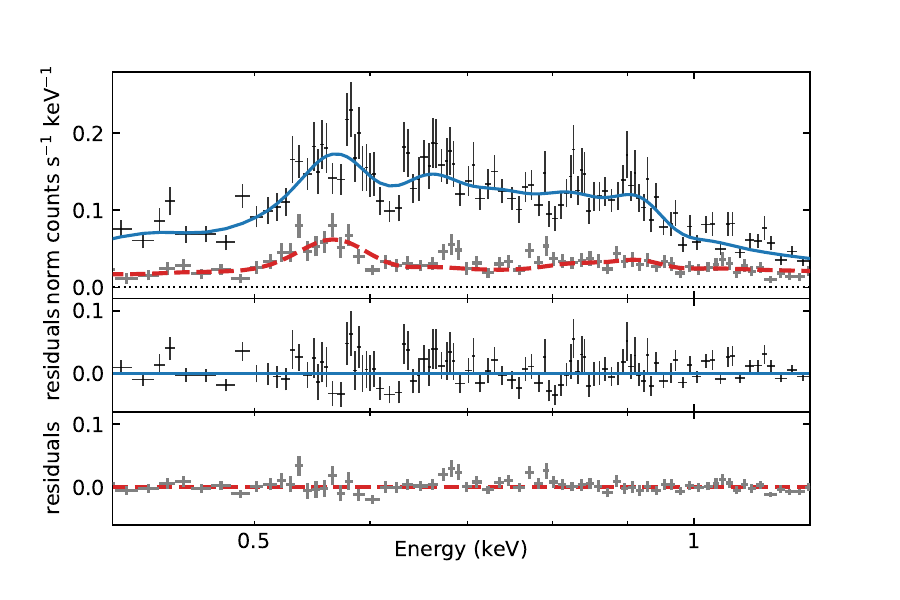}
\vspace*{-0.3 cm}
\caption{\suzaku XIS1 spectra towards MBM36, on-cloud (grey) and off-cloud (black), fitted with the revised model of the SDXB as discussed in \S 3.2.1. The middle (off-cloud) and bottom (on-cloud) panels show the residual plots. }
\end{figure}


\newpage
\begin{deluxetable*}{lcccccccc}
\tabletypesize{\scriptsize}
\tablewidth{0pt} 
\tablenum{1}
\tablecaption{Suzaku Shadow Observations Log \label{tab:deluxesplit}}
\tablehead{
\colhead{Shadow} & \colhead{Target} & \colhead{\it l}& \colhead{\it b} & \colhead{\it Distance$^{a}$} &\colhead{Start Date} & \colhead{Exposure} & \colhead{$\rm N_{H}$$^{b}$} &\colhead{$\rm N_{H}$$^{c}$}\\
\colhead{} & \colhead{} & \colhead{($ ^{\circ}$)} & \colhead{($ ^{\circ}$)} & \colhead{parsec} & \colhead{} & \colhead{(ks)} &  \colhead{$\rm 10^{20}~cm^{-2}$} &  \colhead{$\rm 10^{20}~cm^{-2}$}}
\startdata 
{MBM36} & {On}  & $\rm 3.9$   & $\rm +35.6$ & {$\sim 140$} &{2013 Aug 27} & $\rm 61.8$ & $\rm 24.6$ & $46.2\pm1.7$\\
{}      & {Off} & $\rm 7.4$   & $\rm +37.1$ & {} & {2013 Aug 29} & $\rm 58.7$ & $\rm 6.8$ & \nodata \\
{Filament} & {On}  & $\rm 278.6$ & $\rm -45.3$ & {200-260} & {2006 Mar 03} & $\rm 101.5$ & $\rm 6.7$ & $5.0\pm1.2$ \\
{}      & {Off} & $\rm 278.7$ & $\rm -47.1$ & {} & {2006 Mar 01} & $\rm 80.1$ & $\rm 1.1$ & \nodata \\
{MBM12} & {On}  & $\rm 159.2$ & $\rm -34.5$ & {60-275} &{2006 Feb 03} & $\rm 102.9$ & $\rm 37.3$ & $34.6\pm3.2$\\
{}      & {Off} & $\rm 157.4$ & $\rm -36.8$ & {}  &{2007 Feb 06} & $\rm 75.3$ & $\rm 6.5$ & \nodata\\
{MBM20} & {On}  & $\rm 211.4$ & $\rm -36.6$ & {140-182} &{2008 Feb 11} & $\rm 107.1$ & $\rm 15.7$ & $16.6\pm1.3$ \\
{}      & {Off} & $\rm 213.4$ & $\rm -39.1$ & {} &{2007 Jul 30} & $\rm 103.8$ & $\rm 0.9$ & \nodata\\
{MBM16} & {On}  & $\rm 170.6$ & $\rm -37.3$ & {60-95} &{2013 Aug 07} & $\rm 82.3$ & $\rm 29.5$ & $27.7\pm2.6$\\
{}      & {Off} & $\rm 165.8$ & $\rm -38.4$ & {} &{2013 Aug 09} & $\rm 82.1$ & $\rm 9.3$ & \nodata\\
{G236+38} & {On}  & $\rm 236.0$ & $\rm +38.2$ & {Not Known} & {2011 Jun 01} & $\rm 69.8$ & $\rm 7.4$ & $4.6\pm0.9$ \\
{}      & {Off} & $\rm 237.1$ & $\rm +41.1$ & {} & {2011 Jun 07} & $\rm 70.8$ & $\rm 1.8$  & \nodata\\
\enddata
\tablecomments{\\
$^{a}$ Approximate distances of the molecular clouds\\
$^{b}$ Hydrogen column density estimated using IRAS100$\micron$ data\\
$^{c}$ Best-fit column density for the on-cloud regions. For off-cloud regions, the column density was frozen at the measured values. \\
}
\end{deluxetable*}
https://www.overleaf.com/project/60099b830ab3d9d0f04e2270

\begin{deluxetable*}{lccccccccccccccc}
\tabletypesize{\scriptsize}
\tablewidth{0pt} 
\tablenum{2}
\tablecaption{Best Fit Model Parameters}
\tablehead{
\colhead{Target} & \multicolumn{2}{c}{Foreground} & \colhead{}  & \multicolumn{6}{c}{\bf Galactic Emission} & \colhead{}  &\multicolumn{2}{c}{\bf Power-Law}  &  \colhead{$\chi^{2}/d.o.f.$}\\
\cline{2-3}
\cline{5-10}
\cline{12-13}
\colhead{Model} & \colhead{kT$^{a}$} & \colhead{Norm$^{b}$} & \colhead{}  &  \colhead{kT$_{1}^{c}$} & \colhead{N} & \colhead{Ne} & \colhead{EM$_{1}^{d}$} &\colhead{kT$_{2}^{e}$} &  \colhead{EM$_{2}^{d}$} &  \colhead{}  &\colhead{$\Gamma$} & \colhead{Norm$^{f}$} & \colhead{}
} 
\rotate
\startdata 
{\bf MBM36$^{*}$} & $0.139\pm0.007$ & $0.56\pm0.10$ & & $0.187\pm0.010$ & $1.7\pm0.6$ &  $3.1\pm0.6$ & $13.70\pm1.22$ & $0.454\pm0.084$ & $3.04\pm0.96$ &   & $1.32\pm0.06$  & $9.32\pm0.52$ &$572.65/520$ \\
{\bf Filament$^{\#}$} & \nodata & \nodata & & $0.179\pm0.005$ & $1.6\pm0.3$ &  $2.8\pm0.5$ & $6.59\pm0.50$ & $0.517\pm0.200$ & $0.30\pm0.14$ &   & $1.54\pm0.07$  & $8.86\pm0.39$ &$675.02/672$ \\
{\bf MBM12$^{\$}$} & $0.104\pm0.006$ & $2.91\pm0.85$ & & $0.167\pm0.011$ & $3.1\pm1.3$ &  \nodata & $2.83\pm0.61$ & \nodata  & \nodata &  & $1.30\pm0.06$ & $7.40\pm0.34$ &$545.80/502$  \\
{\bf MBM20} & $0.075\pm0.019$ & $16.60\pm30.80$ & & $0.195\pm0.008$ & $3.7\pm1.2$ &  \nodata & $2.67\pm0.28$ & \nodata  & \nodata &  & $1.44\pm0.04$ & $9.31\pm0.29$ &$642.02/622$  \\
{\bf MBM16} & $0.172\pm0.020$ & $0.54\pm0.11$ & & $0.259\pm0.038$ & \nodata &  \nodata & $1.63\pm0.68$ & \nodata  & \nodata &  & $1.35\pm0.05$ & $9.43\pm0.44$ &$395.18/421$  \\
{\bf G236$+$38} & $0.139\pm0.013$ & $1.94\pm0.66$ & & $0.190\pm0.018$ & $2.7\pm1.7$ &  \nodata & $3.23\pm1.44$ & \nodata  & \nodata &  & $1.43\pm0.05$ & $10.28\pm0.38$ &$455.78/453$  \\
\enddata
\tablecomments{\\
$^{a}$ Temperature of ACX2 (SWCX) component in units of keV.\\
$^{b}$ Normalization of ACX2 component in units of $\rm 10^{-4} ~cm^{-5}$.\\
$^{c}$ Temperature of Galactic halo warm-hot phase thermal component in units of keV.\\
$^{d}$ Emission measure in Units of $\rm 10^{-3}cm^{-6}pc$.\\
$^{e}$ Temperature of Galactic halo hot phase thermal component in units of keV.\\
$^{f}$ Normalization of the Power-law model in the units of $\rm photons~keV^{-1}~s^{-1}~sr^{-1}~cm^{-2}$.\\
$^{*}$ The sightline of \mbmxxxvi off-cloud also required a Gaussian at $\rm E = 0.42 \pm 0.01~keV$, with a line strength of $\rm 4.6 \pm 1.2~LU$. For the on-cloud observation, the line strength was consistent with zero.}
$^{\#}$ For the sightline of the filament in the SGH, the foreground component was modeled using only an APEC model with the temperature fixed at $0.1~\mathrm{keV}$. The EMs were allowed to vary between the on- and off-filament regions, yielding values of $3.20 \pm 1.86$ and $16.1 \pm 2.30~\rm 10^{-3}cm^{-6}pc$, respectively.\\
$^{\$}$ The sightline of \mbmxii off-cloud also required a Gaussian at $\rm E = 0.86 \pm 0.02~keV$, with a line strength of $\rm 0.25 \pm 0.11~LU$. For the on-cloud observation, the line strength was consistent with zero.
\end{deluxetable*}

\clearpage

\bibliographystyle{aasjournal}

\end{document}